\documentclass{singlecol-new}

\usepackage[numbers]{natbib}
\usepackage{stfloats}
\usepackage{mathrsfs}
\usepackage{graphicx}

\theoremstyle{TH}{

}

\theoremstyle{THrm}{

}

\theoremstyle{THhit}{

}

\makeatletter
\def\theequation{\arabic{equation}}

\makeatother

\begin{document}%

\setcounter{page}{1}


\RRH{Geometrical Tools for Teaching Azeotropy}

\VOL{x}

\ISSUE{x}

\PUBYEAR{201X}

\BottomCatch

\CLline

\subtitle{}

\title{Geometrical Tools for Teaching Azeotropy Using Simplified Thermodynamic Models}

%
%
%
%

%
%

%
%
%
%
%
%

\begin{abstract}
In this work we propose a geometric view of the azeotropy problem, using some simplified models. We demonstrate that the occurrence of azeotropes in binary mixtures can be viewed---geometrically---as the intersection of curves in the plane (for some models, these curves are parabolas). Furthermore, the idea of functions from the plane to the plane is used to help understand the azeotropic phenomenon. These ideas are illustrated with two simple cases, with one and two azeotropes, allowing the analysis of a unusual thermodynamic behavior---such as double azeotropy---with simple mathematical tools, by undergraduate students in Chemical Engineering courses.
\end{abstract}

\KEYWORD{Azeotropes; Functions from the Plane to the Plane; Double Azeotropy.}

\maketitle

\section{Introduction}

By definition, an azeotrope is a liquid mixture that produces a vapor with equal relative quantities of each component. Thus, the vapor phase is not ``enriched'' with respect to the more volatile compounds, as might be expected. The existence of the azeotropy phenomenon may introduce difficulties to the separation of such mixtures by distillation. The existence of the azeotrope explains, for example, the common use of the mixture of ethanol and water 96$^{\circ}$ GL (hydrated, hydrous or azeotropic ethanol; GL refers to the Gay-Lussac scale, employed to measure the concentration of ethanol) for pharmaceutical applications, instead of pure ethanol. An azeotrope between ethanol and water prevents the production of pure ethanol (used, for instance, blended with gasoline in Brazil to produce E20-E25 fuel) by ordinary distillation processes.

Consequently, the understanding of azeotropic phenomenon is an essential pre-requisite for the study of enhanced distillation processes \cite{Calvo2016}, an extremely important subject in modern Chemical Engineering curricula. The concepts of residue curve maps, region boundaries and separatrices, for instance, in nonideal systems demand the comprehension of the azeotropy \cite{Doherty1979}. Furthermore, as pointed by Kurtyka \cite{Kurtyka1975}, the existence of azeotropes is not rare (the double azeotropy in binary systems, on the other hand, is extremely unusual), which justifies a comprehensive analysis of the azeotropy conditions.

Some authors discussed---using thermodynamic and algebraic considerations---the conditions for azeotropy in binary systems, illustrating the capabilities of some models to predict the existence
of  azeotropes (and even multiple azeotropes). See, 
for example, Refs. \cite{Wisniak1996} and \cite{Missen2005}. 
The effect of pressure in the azeotropic condition was also analyzed by some authors \cite{Joffe1955, Ruether1972}. Missen \cite{Missen2006} described an algebraic method devoted to the calculation of azeotropic coordinates for some common excess Gibbs free energy models. Guedes et al. \cite{Canadian2015} analyzed---with a geometric approach, using the concept of functions from the plane to the plane---the existence of double azeotropy in two binary systems. Libotte et al. \cite{Industrial2018} used geometrical concepts to explain the double retrograde vaporization phenomenon (another rare and nonlinear thermodynamic phenomenon, which can occur in mixtures close to the critical point).

The study of azeotropy phenomenon, as discussed here, allows 
a better understanding of the main concepts of maps in the plane 
(useful, for instance, in the solution of systems of nonlinear equations) by undergraduate students, as well as illustrates the influence of important aspects in the azeotrope existence, such as system pressure, relative volatility of the components and non-ideality of the liquid phase. 

The main objective of this work is to illustrate that, even using simple models, some interesting thermodynamic behavior---such as azeotropy and double azeotropy---can be predicted and analyzed using a geometrical approach. Furthermore, the persistence of azeotropes can also be understood using this methodology. We also detail the azeotropic conditions using analytic geometry procedures and numerical calculus, indicating a fruitful interaction between these disciplines. In this scenario, for instance, Cogswell \cite{Cogswell2009} analyzed the existence of the azeotropy phenomenon in the mixture methanol-acetone using the calculation of extrema, showing a clear application of the differential calculus techniques in thermodynamics, for undergraduate students (at the University of South Florida, USA).

The focus of this work is essentially different from that employed by Guedes et al. \cite{Canadian2015}, which used some complex mathematical tools (such as homotopy-continuation methods) and highly nonlinear thermodynamic models, making it technically more difficult to demonstrate in elementary terms the basic geometric ideas involved in the azeotropy calculations. The azeotropy phenomenon in binary mixtures can be characterized by a nonlinear application of $ \mathbb{R}^{2} $ to $ \mathbb{R}^{2} $, which motivates us to present some geometric notions on quadratic maps in the plane in Section 2. In Section 3, different ways of modeling the double azeotropy problem are shown. However, regardless of how it is modeled, when the mixture is composed of two components, the problem is always described by a $ 2 \times 2 $ system of nonlinear algebraic equations. Finally, the results are discussed in Section 4 and in Section 5 the conclusions are presented.

\section{Some notions on the geometry of quadratic maps in the plane}

A general treatment of the geometry of nonlinear functions from the 
plane to the plane can be seen in Ref. \cite{tomei}.

\subsection{Quadratic maps and critical sets}

In  general, a quadratic map from the plane 
to the plane can be represented as \cite{Niem2016}:
\begin{eqnarray}
F(x,y) = \left( a_0 x^2 + a_1 xy + a_2 y^2 + a_3 x + a_4 y + a_5, \right. \nonumber \\
\left. b_0 x^2 + b_1 xy + b_2 y^2 + b_3 x + b_4 y + b_5 \right)
\label{eq:map}
\end{eqnarray}
{\em i.e.}, as a pair of second degree polynomials in two variables. The function $F = \left( f_1, \; f_2 \right)$ is a function from a subset 
of $\mathbb{R}^2$  to $\mathbb{R}^2$, that we simply call {\em function
from  the plane to the plane}.

The {\em critical curve} $C$ is defined by $C = \{(x,y) \in \mathbb{R}^2 | \det J = 0 \}$, where we recall that the Jacobian matrix $J$ of the nonlinear application is 
\begin{equation} \label{eq:jacobiana}
J = \nabla F =  
\begin{bmatrix}
    \frac{\partial f_1}{\partial x} & \frac{\partial f_1}{\partial y} \\
    \frac{\partial f_2}{\partial x} & \frac{\partial f_2}{\partial y}
\end{bmatrix}\:.
\end{equation}
\noindent In turn, the image $ F(C) $ is called the \textit{critical image} or \textit{critical locus}.

Considering these definitions, it is intended to show that it is possible to formulate the problem of calculating azeotropes in a binary system through a quadratic map. First, consider the two examples shown below.

\subsection{A simple example}
Some features  in quadratic maps in the plane can be seen by analysing
the function $F=(f_1(x,y),f_2(x,y))=((x+1)(x-5),y)$. Function  
$F$ maps $\mathbb{R}^2$ onto the semiplane $x\geq -9$. The Jacobian of $F$ is
\begin{eqnarray}
J=\left[
\begin{array}{cc} 2x-4 & 0 \\ 0 & 1
\end{array}
\right] \; .
\end{eqnarray}
Thus the critical curve is $x=2$, 
and the critical image is $x=-9$. The $y$ variable and the function 
$f_2$
 do not play a relevant role in this example, and what happens is that
 the plane is folded in two pieces, `bending' and `stretching' 
 along the $x$ axis, with {\em folding line} $x=2$ (the critical 
 curve),
  which is mapped into the line $x=-9$ (the critical image).
   All points in the codomain that have 
  $x$ coordinate greater than $-9$ 
 have two pre-images (that is, the equation $(x+1)(x-5)=b$ has two solutions, if $b>-9$), for points that have 
  $x$ coordinate less than $-9$ there are no pre-images (if $b<-9$, then equation $(x+1)(x-5)=b$ has no solutions), and a point in the
  transition curve $x=-9$ has just one pre-image, 
 with $x=2$, and whatever $y$ coordinate it had.
  In the case that there are two pre-images, one is on one side of the critical curve and the other is on the other side. For instance, the point  $(7,3)$ with
 $b=7$ has two pre-images, $(x,y)=(-2,3)$ and $(x,y)=(6,3)$,
  one to the left
  and the other to the right of  the 
 critical curve given by $x=2$.

In fact, this example implicitly represents a case of a quadratic map from the line to the line, which share
 some features with more general quadratic maps in the plane.

\subsection{A more interesting example}

We will consider, as an example, the following quadratic map:
\begin{eqnarray}
F(x,y) = \left( -x^2 + x + y + 1, -x^2 +2y + 1 \right)
\label{eq:map_coef} 
\end{eqnarray}

The critical curve is the line $x=1$ and the critical image is represented as $F(1,y)=(1+y,2y)$, for all $y\in \mathbb{R}$.

Now, we consider
\begin{eqnarray}
\bar{F}(x,y) = \left( -x^2 + x + y, -x^2 +2y \right)
\label{eq:map_coef_form2}
\end{eqnarray}

It is clear that ${F}(x,y) = (0,0)$ is equivalent to $\bar{F}(x,y) = (-1,-1)$, or $\bar{F}(x,y) = q$ with $q=(-1,-1)$. Furthermore, the critical curves for functions $F$ and $\bar{F}$ are the same, but the critical images are different.

\begin{figure*}[t] \center 
\caption{(a) A disk with radius 1 and the critical curve. (b) Images of the disk by $F$ (continuous line) and $\bar{F}$ (dashed line) and the critical images of $F$ (continuous line) and $\bar{F}$ (dashed line). Also in (b), the cross $\times$  represents the point $q=(-1,-1)$. Note that only the right-hand side of $F$ and $\bar{F}$ are different, and yet the functions describing the critical curve of both is the same.} \label{fig:quadratic_map}
\centerline{\includegraphics[width=0.9\textwidth]{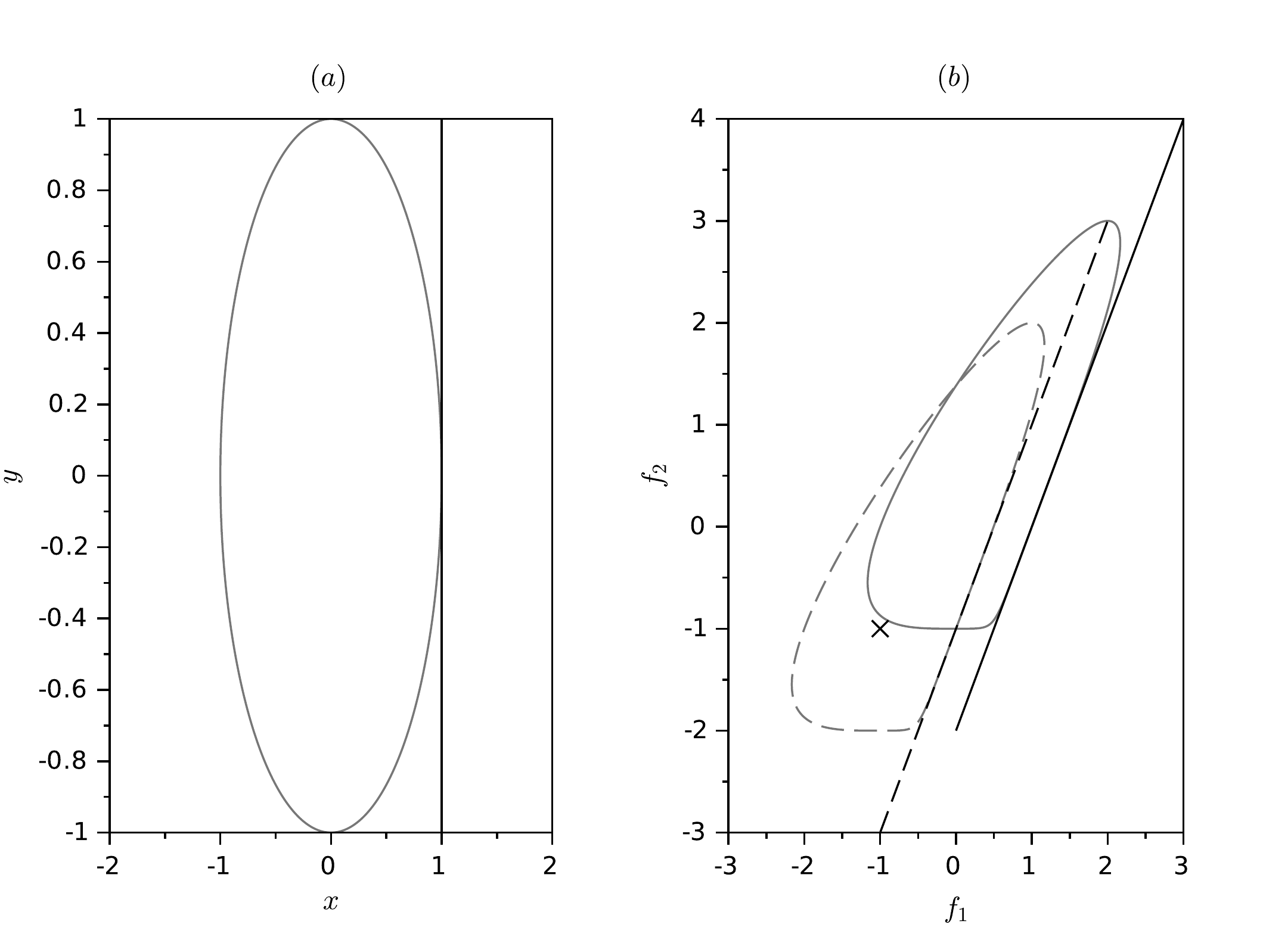}}
\end{figure*}

Figure \ref{fig:quadratic_map}a contains a disk with radius 1 and the critical curve, while  Fig.~\ref{fig:quadratic_map}b exhibits 
the images of the disk with radius 1, and the critical images,  for maps $F$ (continuous line)  and $\bar{F}$ (dashed line). 

The nonlinear algebraic system $\bar{F}(x,y)=q$ can be represented as:
\begin{subequations}
\begin{eqnarray}
-x^2 + x + y - q_1 = 0 \\
-x^2 +2y - q_2 = 0
\end{eqnarray}
\end{subequations}
which, for a given $q$, corresponds to the intersection of two
parabolas. 
Thus $y -x = q_2 - q_1$, and then:
\begin{equation}
-x^2 + x + x + q_2 - q_1 - q_1 = 0 \rightarrow x^2 - 2x + 2q_1 - q_2=0
\end{equation}
The  solutions of the system are:
\begin{eqnarray}
x =  1 \pm \sqrt{1 - 2q_1 + q_2}\;, \qquad & & y=1+ q_2-q_1
 \pm \sqrt{1 - 2q_1 + q_2}
 \label{50}
\end{eqnarray}

For instance, from Eqs.~(\ref{50}), the point $(-1,-1)$ in the image 
of $\bar{F}$, represented by a cross in Fig.~\ref{fig:quadratic_map}b,
has two pre-images (in the domain),
\begin{eqnarray*}
(x_{\pm},y_{\pm})=(1\pm \sqrt{2},1\pm \sqrt{2})
\end{eqnarray*}


\subsection{Collision of pre-images}
\begin{figure*}[t] \center 
\caption{(a) Path of the pre-images colliding in the critical curve
 (domain) (b) Movement towards the critical image maintaining $q_2 = -1$. When the critical image is traversed, the sequence of points in the domain (pre-images) collide at exactly the same point of the critical curve, causing degeneration.} \label{fig:degenera_parab}
\centerline{\includegraphics[width=0.9\textwidth]{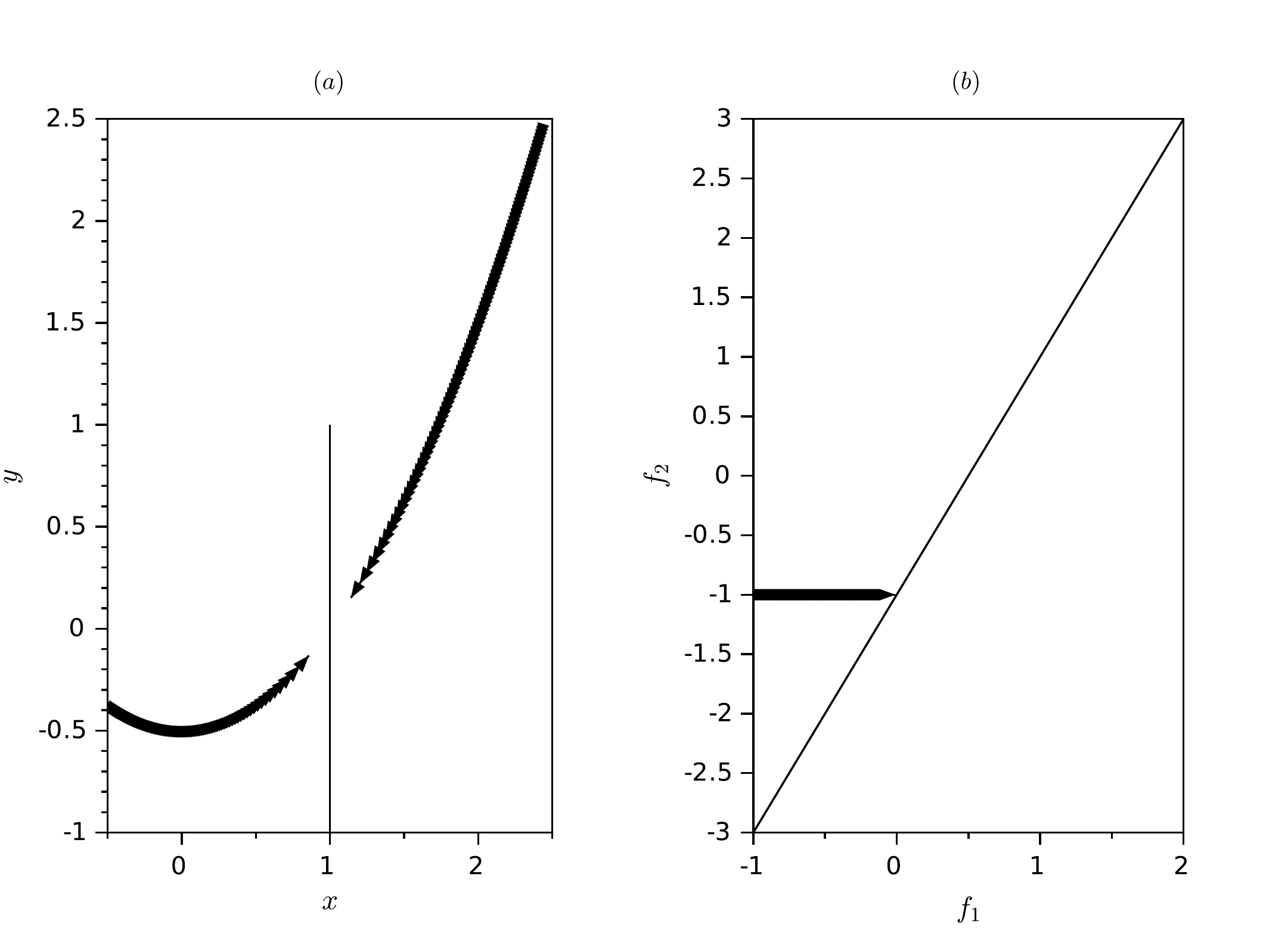}}
\end{figure*}

An interesting situation occurs if we consider a change in the value 
of $q$, beginning in $q=(-1,-1)$ moving in the direction of the 
critical image, 
maintaining $q_2=-1$ constant and increasing $q_1$
  from $q_1 = -1$ with equally spaced steps of length $0.1$. 
 Figure \ref{fig:degenera_parab} illustrates the behavior 
 of the pre-images (in the domain) as the value of $q_1$  
 increases. Clearly, we observe that the two pre-images
  degenerate at $q_1$, {\em i.e.}, the two intersections 
  of the parabolas become only one (collide). In other words, the 
  region to the left of the critical image has 2 pre-images 
  and the portion to the right shows no pre-images. This 
  behavior is consistent with a \textit{fold} of the domain over the
  codomain with the critical curve as the folding locus in the domain,
   which has the
  critical image as its image---the folding locus in the codomain, a transition curve---comprising the boundary of the regions where
  the points have two pre-images and no pre-image. 


\section{Problem Description}

Classically, the azeotrope calculation in a binary system modeled by the modified Raoult's law is represented by the following nonlinear algebraic system \cite{Walas}:
\begin{subequations}
\label{eq:azeotropy}
\begin{eqnarray}
P &= \gamma_1 P_1^{sat}  \\
P &= \gamma_2 P_2^{sat} 
\end{eqnarray}
\end{subequations}
where $P$ is the system pressure, $\gamma_i$ refers to the activity coefficient for component $i$, and $P_i^{sat}$ is the respective 
saturation pressure of the pure component. 

Eqs. (\ref{eq:azeotropy}) can be restated as:
\begin{subequations}
\label{eq:azeotropy_log}
\begin{eqnarray}
\ln P - \ln \gamma_1 - \ln P_1^{sat}= 0  \\
\ln P - \ln \gamma_2 - \ln P_2^{sat} =0
\end{eqnarray}
\end{subequations}

We will present two different models leading to the  azeotropy 
phenomena, using a symmetrical Margules model (Model A), and a Margules model with
two distinct binary interaction parameters (Model B), for the activity coefficient. Furthermore, two formulations for the nonlinear algebraic problem (Formulations 1 and 2) will be discussed. 

In Formulation 1 we are interested in solving\\

\begin{equation}
F(p) = 0\:, 
\end{equation}

\noindent \\to find the azeotropes, 
where $p \in \mathbb{R}^{2}$---characterizing the azeotrope---is a point in the domain (the coordinates of the vector $p$ 
are dependent on the type of the problem---isobaric or isothermal),
 $0$ is the null vector in the $\mathbb{R}^2$ plane and therefore $F$ is a function from the plane to the plane.
 Function $F$ needs not to be defined in the whole plane due to non-physical values or 
singularities in its definition.

On the other hand, in Formulation 2, we solve
\begin{equation}
F(p) = q\:,
\end{equation}
where $q$ is a non-null vector in the codomain, which varies,
with the aim of
analysing the dependence of the azeotropes on the pressure of the
system or on 
parameters related to the vapor pressure of  pure components.

Moreover, for each problem we still 
have the possibility to solve isobaric or isothermal problems. 
We present them next.

\subsection{Model A (Problem 1A)}

In  Model A, we  use the symmetrical Margules model (or Porter model) for the activity coefficient, as follows \cite{Walas}:
\begin{subequations}
\label{eq:margules}
\begin{eqnarray}
\ln \gamma_1 &=& A (1-x_1)^2  \\
\ln \gamma_2 &=& A x_1^2
\end{eqnarray}
\end{subequations}
where $x_1$ is the molar fraction in the liquid phase of component 1. Here, only one parameter ($A$) is employed to describe the nonideality of the liquid phase. We also consider that $A$ is independent of temperature and pressure (anyway, the influence of pressure in the non-ideality of the liquid phase is usually small).

Moreover, a simplified model is employed for the calculation of saturation pressures, also called August equation \cite{Altschuh2015}:
\begin{subequations}
\label{eq:saturation}
\begin{eqnarray}
\ln P_1^{sat} = B_1 + \frac{C_1}{T}  \\
\ln P_2^{sat} = B_2 + \frac{C_2}{T}
\end{eqnarray}
\end{subequations}
where $T$ represents the system temperature. The terms
 $B_1$ and $B_2$ are constants, which depend on pure substances.

In fact, vapor pressure of pure compound $i$ can be 
represented as \cite{Rathbun1987}:
\begin{equation}
\ln P_i^{sat} = 17.23 - 10.6 \frac{T_{b,i}}{T}
\label{eq:psat}
\end{equation}
where $P_i^{sat}$ is in mmHg and $T$ in Kelvin. $T_{b,i}$ is the normal boiling point temperature (also in Kelvin). Then, we 
 note that (using this model) $B_1 = B_2 = B$, and $C_i = -10.6 T_{b,i}$.

\subsubsection*{Isobaric systems}

In an isobaric process, the value of $P$ is known. With 
Eqs. (\ref{eq:margules}) and (\ref{eq:saturation}) 
in Eqs. (\ref{eq:azeotropy_log}):
\begin{subequations}
\label{eq:azeotropy_final}
\begin{eqnarray}
\ln P - A (1-x_1)^2 - B_1 - \frac{C_1}{T} = 0 \\
\ln P - A x_1^2 - B_2 - \frac{C_2}{T} = 0
\end{eqnarray}
\end{subequations}
Considering $T^{\ast} = \frac{1}{T}$:
\begin{subequations}
\label{eq:azeotropy_final_linear}
\begin{eqnarray}
\ln P - A (1-x_1)^2 - B_1 - C_1 T^{\ast} &= &0  \\
\ln P - A x_1^2 - B_2 - C_2 T^{\ast} &=& 0
\end{eqnarray}
\end{subequations}
Finally, the problem can be interpreted as 
root finding, $F=0$, of 
a function from the plane to the plane, $F=(f_1,f_2)$, with components:
\begin{subequations}
\label{eq:azeotropy_log_fun}
\begin{eqnarray}
f_1\left(x_1,T^{\ast}\right) 
&=& \ln P - A (1-x_1)^2 - B_1 - C_1 T^{\ast}  \\
f_2\left(x_1,T^{\ast}\right)& = &\ln P - A x_1^2 - B_2 - C_2 T^{\ast}
\end{eqnarray}
\end{subequations}

\noindent A comparison between Eqs. (\ref{eq:map}) and (\ref{eq:azeotropy_log_fun}) indicates that these azeotropy problem calculations can be described by  quadratic maps in the plane, with coefficients 
$a_1$, $a_2$, $b_1$, $b_2$ and $b_3$ null. One may also notice that the 
positive coefficient for $y$ in Eq. (\ref{eq:map_coef}) is consistent with the negative values for $C_1$ and $C_2$ in Eqs.~(\ref{eq:azeotropy_log_fun}).

Since we are interested in the case where $f_1 = 0$ and $f_2 = 0$,
 under specified pressure, by solving each of these equations for $T^*$ as a function of $x_1$, 
 the azeotropy calculation problem can be 
 treated as the intersection of two parabolas in the plane, {\em i.e.}:
\begin{subequations}
\label{eq:azeotropy_log_final}
\begin{eqnarray}
T^{\ast}& = &\frac{\ln P-B_1}{C_1} - \frac{A}{C_1} (1-x_1)^2  \\
T^{\ast} &= &\frac{\ln P-B_2}{C_2} - \frac{A}{C_2} x_1^2 
\end{eqnarray}
\end{subequations}

Clearly, the number of intersection points of the two parabolas  depends on the parameters $A$, $B_1$, $B_2$, $C_1$, $C_2$ and $\ln P$. By the Bezout's theorem \cite{Bizu}, the number of intersections of two conicals in the plane is four (counting multiplicities and allowing 
complex coordinates; here we are interested only in solutions with real coordinates).

\subsubsection*{Critical curve and critical image}
Now we illustrate how to obtain the critical curve for the
nonlinear function $F$,  defined by Eqs.~(\ref{eq:azeotropy_log_fun}). Consider the definition given by Eq. (\ref{eq:jacobiana}). In this case, the critical curve is given by $C = \{(x_1,T^{*}) \in \mathbb{R}^2 | \det J = 0 \}$, where
\begin{equation}
J = 
\begin{bmatrix}
    2A(1-x_1) & -C_1 \\
    -2Ax_1 & -C_2
\end{bmatrix}\:. 
\end{equation}
Thus $\det J = 0$ only if
\begin{equation}
x_1 = \frac{C_2}{C_2-C_1}
\label{eq:critical_curve_1A}
\end{equation}

Therefore, there is only one critical curve which is a vertical line
in the plane. The critical image,
 the image of the critical curve  under $F$, 
 is obtained by plugging Eq.~(\ref{eq:critical_curve_1A}) in Eq.~(\ref{eq:azeotropy_log_fun}), which, due to the arbitrariness of $T^*$, is a line through $\left(
 \ln P - A \left(\frac{C_1}{C_2-C_1}\right)^2-B_1, 
 \ln P - A \left(\frac{C_2}{C_2-C_1}\right)^2-B_2\right)$ and parallel
 to $(C_1,C_2)$.

 An interesting fact regarding Eq.~(\ref{eq:critical_curve_1A}) is that the critical curve never appears in the 
feasible domain ($0 < x_1 < 1$).
 In fact, considering Eq.~(\ref{eq:psat}), it is clear that the physical problem imposes that $C_j<0$, since absolute temperatures satisfy $T_{b,i}>0$. If $\frac{C_2}{C_2-C_1}>0$ and $C_2<0$, then $C_2-C_1<0 $, which implies that $ C_2<C_1$. On the other hand, with $C_2<0$ and $C_2-C_1<0$, the ratio $\frac{C_2}{C_2-C_1}<1$ implies that $C_2>C_2-C_1$ and, then, $C_1>0$, which violates the premise $C_1<0$.

\subsubsection*{Isothermal systems}

For isothermal systems, $P_1^{sat}$ and $P_2^{sat}$ are known. Furthermore, we will consider the following transformed variable $P^{\ast} = \ln P$. By algebraic manipulations, similar to those previously performed, we obtain:
\begin{subequations}
\label{eq:azeotropy_log_fun_isot}
\begin{eqnarray}
f_1\left(x_1,P^{\ast} \right) &=& P^{\ast} - A (1-x_1)^2 - \ln P_1^{sat} \\
f_2\left(x_1,P^{\ast} \right) &= &P^{\ast} - A x_1^2 - \ln P_2^{sat}
\end{eqnarray}
\end{subequations}
Once again we are looking for $f_1 = 0$ and $f_2 = 0$. In this case, only one real solution exists for the azeotropic composition:
\begin{equation}
x_1 = \frac{1}{2}\left( \tau + 1 \right)
\end{equation}
where $\tau = \frac{1}{A}\ln \left( \frac{P_1^{sat}}{P_2^{sat}} \right)$. This familiar result was obtained, for instance, by Refs. \cite{Walas}
and \cite{Eubank2004}.

The problem can still be seen as the intersection of two parabolas. But, in this particular case, the two parabolas have the same maximum point and exhibit only one intersection point.

The deduction of the critical curves and critical images could be made for the isothermal case; but the results will be focused on the isobaric situation. For this reason, the critical curve and critical image when $T$ is fixed will not be detailed.

\subsection{Model B (Problem 1B)}
	
Now we consider the model B, using the Margules model with two distinct binary interaction parameters for the activity 
coefficient \cite{Walas}
\begin{subequations}
\label{eq:M2A}
\begin{eqnarray}
\ln \gamma_{1}&=&(1-x_{1})^{2}\left[A_{12}+2(A_{21}-A_{12})x_{1}\right]
 \\
\ln \gamma_{2}&=&x_{1}^{2}\left[A_{21}+2(A_{12}-A_{21})(1-x_{1})\right]
\end{eqnarray}			   
\end{subequations}

The expressions for saturation pressures are the same as in Model A, August equation~(\ref{eq:saturation}).

\subsubsection*{Isobaric systems}
		
With Eqs. (\ref{eq:M2A}) and (\ref{eq:saturation}) in Eqs. (\ref{eq:azeotropy_log}), we obtain:
\begin{subequations}
\begin{eqnarray}
\ln P-(1-x_{1})^{2}[A_{12}+2(A_{21}-A_{12})x_{1}]-B_1-\frac{C_{1}}{T}&=&0
\\
\ln P-x_{1}^{2}[A_{21}+2(A_{12}-A_{21})(1-x_{1})]-B_2-\frac{C_{2}}{T}&=&0
\end{eqnarray}
\end{subequations}
We continue considering $T^{*} = \frac{1}{T}$, and thus we have
\begin{subequations}
\label{19}
\begin{eqnarray}
\ln P-(1-x_{1})^{2}[A_{12}+2(A_{21}-A_{12})x_{1}]-B_1-C_{1}T^{*}&=&0
\\
\ln P -x_{1}^{2}[A_{21}+2(A_{12}-A_{21})(1-x_{1})]-B_2-C_{2}T^{*}&=&0
\end{eqnarray}
\end{subequations}

This leads to the definition of 
 an application from the plane to the plane, $F=(f_1,f_2)$,
with:
\begin{small}
\begin{subequations}
\label{equa}
\begin{eqnarray}
f_{1}(x_1,T^{*})&=&\ln P -(1-x_{1})^{2}[A_{12}+2(A_{21}-A_{12})x_{1}]-B_1-C_{1}T^{*}
\\
f_{2}(x_1,T^{*})&=&\ln P -x_{1}^{2}[A_{21}+2(A_{12}-A_{21})(1-x_{1})]-B_2-C_{2}T^{*}
\end{eqnarray}
\end{subequations}
	\end{small}

Once more, under specified pressure, the problem of calculating azeotropy can be seen as the intersection of curves
 in the plane, {\em i.e.}: 		
\begin{subequations}
\begin{eqnarray}
T^{*}&=&\frac{\ln P}{C_{1}} -\frac{(1-x_{1})^{2}}{C_{1}}[A_{12}+2(A_{21}-A_{12})x_{1}] - \frac{B_1}{C_{1}}
\\
T^{*}&=&\frac{\ln P}{C_{2}} -\frac{x_{1}^{2}}{C_{2}}[A_{21}+2(A_{12}-A_{21})(1-x_{1})] - \frac{B_2}{C_{2}}
\end{eqnarray}
\end{subequations}
In this case, however, the curves no longer represent parabolas, but can be rewritten as {\em cubic curves}, {\em i.e.} the graph of cubic polynomials in $x_1$:
\begin{small}	
\begin{subequations}
\label{22}
\begin{eqnarray}
\lefteqn{T^{*}= -\frac{1}{C_{1}}
\left[2(A_{21}-A_{12})x_{1}^{3} + (5A_{12} - 4A_{21})x_{1}^{2}\right.}
\nonumber
\\ && \left. + 2(A_{21} - 2A_{12})x_{1}\right]
+\frac{\ln P -A_{12}-B_1}{C_{1}}
\\
T^{*}&=& -\frac{1}{C_{2}}\left[2(A_{21} -A_{12})x_{1}^{3} + (2A_{12}-A_{21})x_{1}^{2}\right]+\frac{\ln P -B_2}{C_{2}}
\end{eqnarray}
\end{subequations}
\end{small}

\subsubsection*{Critical curves}	
Similarly to the previous formulation, we 
 obtain the critical curve, with the Jacobian of $F$, Eqs.~(\ref{equa}):
 \begin{small}
\begin{equation}	
J=\left[
\begin{array}{c@{\qquad}c}
-6(A_{21}-A_{12})x_{1}^{2} -2(5A_{12}-4A_{21})x_{1} -2(A_{21} -2A_{12}) & -C_{1} \\
6(A_{12} -A_{21})x_{1}^{2} - 2(2A_{12} - A_{21})x_{1}   & -C_{2} 
\end{array}
\right]
\end{equation}
\end{small}

Two $x_{1}$ values satisfy the equation for $\det J=0$,
 accordingly to $x_{1} = \frac{-b \pm \sqrt{b^{2}-4ac}}{2a}$, with
 \begin{subequations}
\begin{eqnarray}
a&=&6\left(\frac{C_{2}}{C_{1}} - 1\right)(A_{12}-A_{21})
\\
b&=&2\left[\frac{C_{2}}{C_{1}}(4A_{21}-5A_{12})-(A_{21}-2A_{12})\right]
\\
c&=& 2\left[\frac{C_{2}}{C_{1}} (2A_{12}-A_{21})\right]
\end{eqnarray}
\end{subequations}

Therefore, we have two critical curves (parallel to $T^{\ast}$ axis), with one critical curve in the physical domain.
		
\subsubsection*{Isothermal systems}
		
From Eqs. (\ref{eq:azeotropy_log}) and (\ref{eq:M2A}), considering the transformed variable $P^{*} = \ln P$ and with fixed values for $P^{sat}_{1}$ and $P^{sat}_{2}$ , we have:
\begin{subequations}
\begin{eqnarray}
f_{1}(x_1,P^{*})&=&P^{*}-(1-x_{1})^{2}[A_{12}+2(A_{21}-A_{12})x_{1}]-\ln P^{sat}_{1}
\\
f_{2}(x_1,P^{*})&=&P^{*}-x_{1}^{2}[A_{21}+2(A_{12}-A_{21})(1-x_{1})]-\ln P^{sat}_{2}
\end{eqnarray}
\end{subequations}
		
Unlike Model A, in this case two solutions appear (considering that $f_1$ and $f_2$ are null and solving the equations for $P^{*}$):
\begin{subequations}
\begin{equation}
x_{1} = \frac{-b \pm \sqrt{b^{2}-4ac}}{2a}
\end{equation}
with
\begin{equation}
a=3(A_{21}-A_{12})
\end{equation}
\begin{equation}
b=2(2A_{12}-A_{21})
\end{equation}
\begin{equation}
c=\ln\left(\frac{P^{sat}_{2}}{P^{sat}_{1}}\right) -A_{12}
\end{equation}
\end{subequations}
This result implies that the existence of a double azeotrope can be predicted with Model B, depending on the quantities 
$A_{12}$, $A_{21}$, $P^{sat}_1$ and $P^{sat}_2$.

Again, we are mainly interested in the numerical results for the isobaric case; thus, the critical curve for the isothermal situation will not be presented.

\subsection{Model A (Problem 2A)}

In the Formulation 2 we will consider only isobaric problems (with specified pressure; thus the modelling for isothermal problems will not be presented). 

Formulation 2 is similar to Formulation 1, but considers a nontrivial 
term in the right side of the nonlinear algebraic system. In other 
words, we will write the system as $F(p) = q$. We recall that in Formulation~1, $
q = 0$. Now, in Formulation~2, $q = (-\ln P,-\ln P)$ or 
$q=(B_1,B_2)$. Obviously, the critical curves coincide,
 since the 
system does not 
depend on the pressure or $B_1$ and $B_2$ values. On the other hand, the critical images are different. As pointed out previously, the 
azeotropy problems 1A and 2A  are represented by  quadratic maps in the plane.

This formulation is more useful because the different nonlinear systems solved represent the effect of the pressure or the values of $B_1$ and $B_2$ on the persistence of the azeotrope. This characteristic will be initially explored using an even simpler model, detailed next.

\subsubsection*{Azeotropes dependence on the pressure}

From Eqs.~(\ref{eq:azeotropy_final_linear}) we will consider, in a first approach, $B_1 = B_2 = B$, in the following way,
\begin{subequations}
\label{eq:formul_21}
\begin{eqnarray}
f_1\left(x_1, T^{\ast}\right)& = &- A (1-x_1)^2 - B - C_1 T^{\ast}
\\
f_2\left(x_1, T^{\ast}\right)& =& - A x_1^2 - B - C_2 T^{\ast}
\label{eq:formul_22}
\end{eqnarray}
\end{subequations}
and $q = \left(-\ln P, -\ln P \right)$.

Considering $F=(f_1,f_2)=(-\ln P, - \ln P)$ in 
Eqs. (\ref{eq:formul_21}), and by eliminating $B$, we obtain:
\begin{equation}
T^{\ast} = \frac{A(1-2x_1)}{C_2-C_1}
\label{eq:tstar}
\end{equation}

\noindent With Eq. (\ref{eq:tstar}) in  Eq. (\ref{eq:formul_22}) and considering $f_2=-\ln P$:
\begin{equation}
x_1^2 - \frac{2C_2}{C_2-C_1}x_1 + \frac{B}{A} - \frac{\ln P}{A} + \frac{C_2}{C_2-C_1}=0
\label{eq:quadratic_equation}
\end{equation}

\noindent The compositional coordinates of the azeotropes are then:
\begin{eqnarray}
x_1 = 
\frac{C_2}{C_2-C_1}\pm \sqrt{\left(\frac{C_2}{C_2-C_1}\right)^2-\left( \frac{B}{A} - \frac{\ln P}{A} + \frac{C_2}{C_2-C_1} \right)}
\label{eq:sol_analit}
\end{eqnarray}

\subsubsection*{Azeotropes dependence on parameters of vapor pressure of pure components}
Once more, from Eqs. (\ref{eq:azeotropy_final_linear}),
 we will consider $q = \left( B_1, B_2 \right)$, {\em i.e.}, with different values for the parameter $B$ for each component (the applicability of this formulation will be clarified in the results). In this case, the azeotrope calculation is described by:
\begin{subequations}
\label{eq:formul_21_B}
\begin{eqnarray}
f_1\left(x_1, T^{\ast}\right) &=& \ln P - A (1-x_1)^2 - C_1 T^{\ast}
\\
f_2\left(x_1, T^{\ast}\right)& =& \ln P - A x_1^2 - C_2 T^{\ast}
\label{eq:formul_22_B}
\end{eqnarray}
\end{subequations}
and $ q = \left( B_1, B_2 \right)$.

By solving the system of equations $(f_1,f_2)=q$, 
Eqs.~(\ref{eq:formul_21_B}), for $T^*$ with respect to $x_1$
gives:
\begin{equation}
T^{\ast} = \frac{B_1 - B_2 + A(1-2x_1 )}{C_2 - C_1}
\end{equation}

Again, the azeotropic coordinates can be obtained by a quadratic equation:
\begin{equation}
x_1^2 - \frac{2C_2}{C_2-C_1}x_1 + \frac{1}{A}\left(B_2 - \ln P +
 \frac{C_2\left(B_1 - B_2 +A \right)}{\left(C_2-C_1\right)}\right)=0
\label{eq:sol_analit_B_mod}
\end{equation}
Obviously, if $B_1=B_2$, then Eq. (\ref{eq:sol_analit_B_mod}) becomes  Eq. (\ref{eq:quadratic_equation}).

\subsubsection*{Model B (Problem 2B)}

The analysis of Problem 2B is similar to that detailed for Problem 2A, but with a more complex model for the activity coefficient. Again, we will consider that $F(\theta)=q$, with $q=(-\ln P,-\ln P)$ or $q=(B_1,B_2)$. Denoting $q=(q_1,q_2)$ and $p=(x_1,T^*)$, 
 the nonlinear system for a isobaric problem is described by 
 $F(p)=q$ with
\begin{subequations}
\begin{eqnarray}
f_{1}\left(x_1, T^{\ast}\right) &=& -(1-x_{1})^{2}[A_{12}+2(A_{21}-A_{12})x_{1}]-B-C_{1}T^{*}
\\
f_{2}\left(x_1, T^{\ast}\right)& =& -x_{1}^{2}[A_{21}+2(A_{12}-A_{21})(1-x_{1})]-B-C_{2}T^{*}
\end{eqnarray}
\end{subequations}
with $q=(-\ln P,-\ln P)$, if we want to analyze the effect of pressure
on azeotropes.

Likewise, since we focus on the effect of parameters related to the vapor pressure of pure components on the azeotropes, we let
\begin{subequations}
\begin{eqnarray}
f_{1}\left(x_1, T^{\ast}\right)&=&\ln P-(1-x_{1})^{2}[A_{12}+2(A_{21}-A_{12})x_{1}]-C_{1}T^{*}
\\
f_{2}\left(x_1, T^{\ast}\right)&=&\ln P-x_{1}^{2}[A_{21}+2(A_{12}-A_{21})(1-x_{1})]-C_{2}T^{*}
\end{eqnarray}
\end{subequations}
with $q=(B_1,B_2)$.

In this case, the azeotropic coordinates (molar fractions $x_1$) can be obtained by solving a cubic equation (not detailed here, since we are mainly interested in the geometric behavior of the curves).

%
%

\section{Results and discussion}

In this section we present the results concerning the situations detailed in Section 3. The azeotropic coordinates are obtained using the analytical expressions for the quadratic cases. In the situation where the model is described by a cubic polynomial, a simple Newton-Raphson procedure is applied (but we also illustrate the azeotropic condition as an intersection of the curves in the plane).

\subsection{Problem 1A}

Initially, we will present the vapor-liquid equilibrium diagram for the system ethanol (1) + benzene (2) (under specified pressure) with the simplified approach as modelled by Problem 1A.

The normal boiling temperature for ethanol and benzene are, respectively, $T_{b,1} = 78.37+273.15=351.5$ and 
$T_{b,2} = 80.1+273.15=353.3$ K \cite{Walas}. The coefficient for symmetrical Margules model employed here was $A=1.25$.  Let 
$B_1=B_2 =B= 17.2$, $C_1 = -10.6T_{b,1} = - 3726.112$ and $C_2 = -10.6T_{b,2} = -3744.45$. With these values for the parameters of the 
isobaric Problem 1A, Eqs.~(\ref{eq:azeotropy_final_linear}), the azeotropic behavior can be simulated.

\subsubsection*{Phase diagram}
 
\begin{figure*}[t] \center
\caption{Phase diagram for ethanol (1) + benzene (2) at 760 mmHg, with the occurrence of an azeotrope in the feasible domain.}
\centerline{\includegraphics[width=0.9\textwidth]{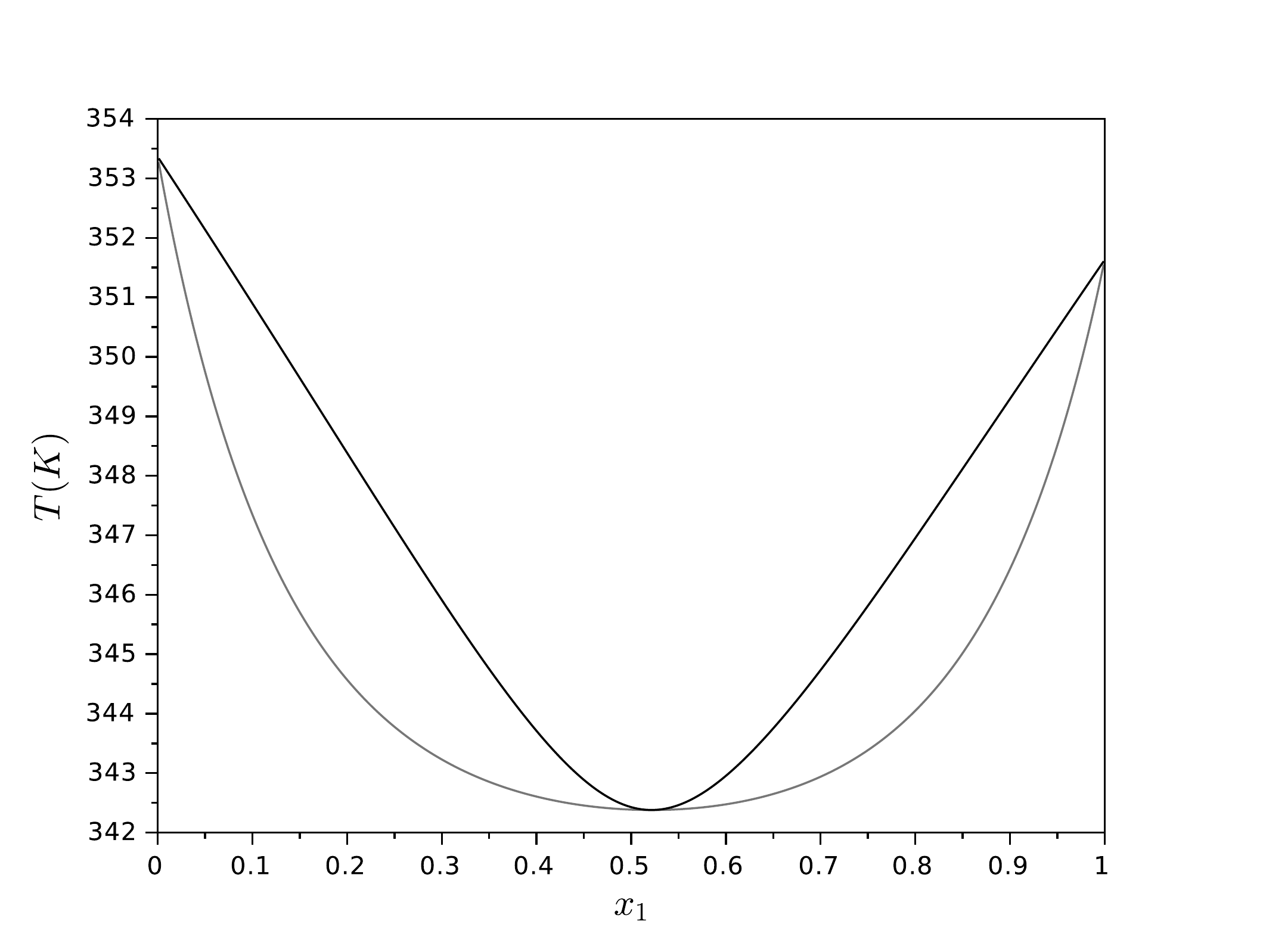}}
\label{fig:phase_diagram}
\end{figure*}

Figure \ref{fig:phase_diagram} contains the isobaric phase diagram (at 760 mmHg) for the binary system ethanol (1) + benzene (2). It is possible to observe the presence of the minimum boiling azeotrope, as expected in the real behavior. However, the computed molar fraction and azeotropic temperature differ from the true experimental value.

\subsubsection*{Plotting the parabolas}

The nonlinear system that represents the azeotrope calculation problem shows two solutions: one with physical significance and the other one with molar fraction $x_1$ greater than one. The two solutions are: (i) $(x_1,T^{\ast})=(0.5213, 0.0029)$ and (ii) $(x_1,T^{\ast})=(407.8601, 55.5349)$.

\begin{figure*}[t] \center
\caption{The intersection of the two parabolas in the physical domain (azeotropic condition). Note the similarity of the molar fraction with Fig. \ref{fig:phase_diagram}.}
\centerline{\includegraphics[width=0.9\textwidth]{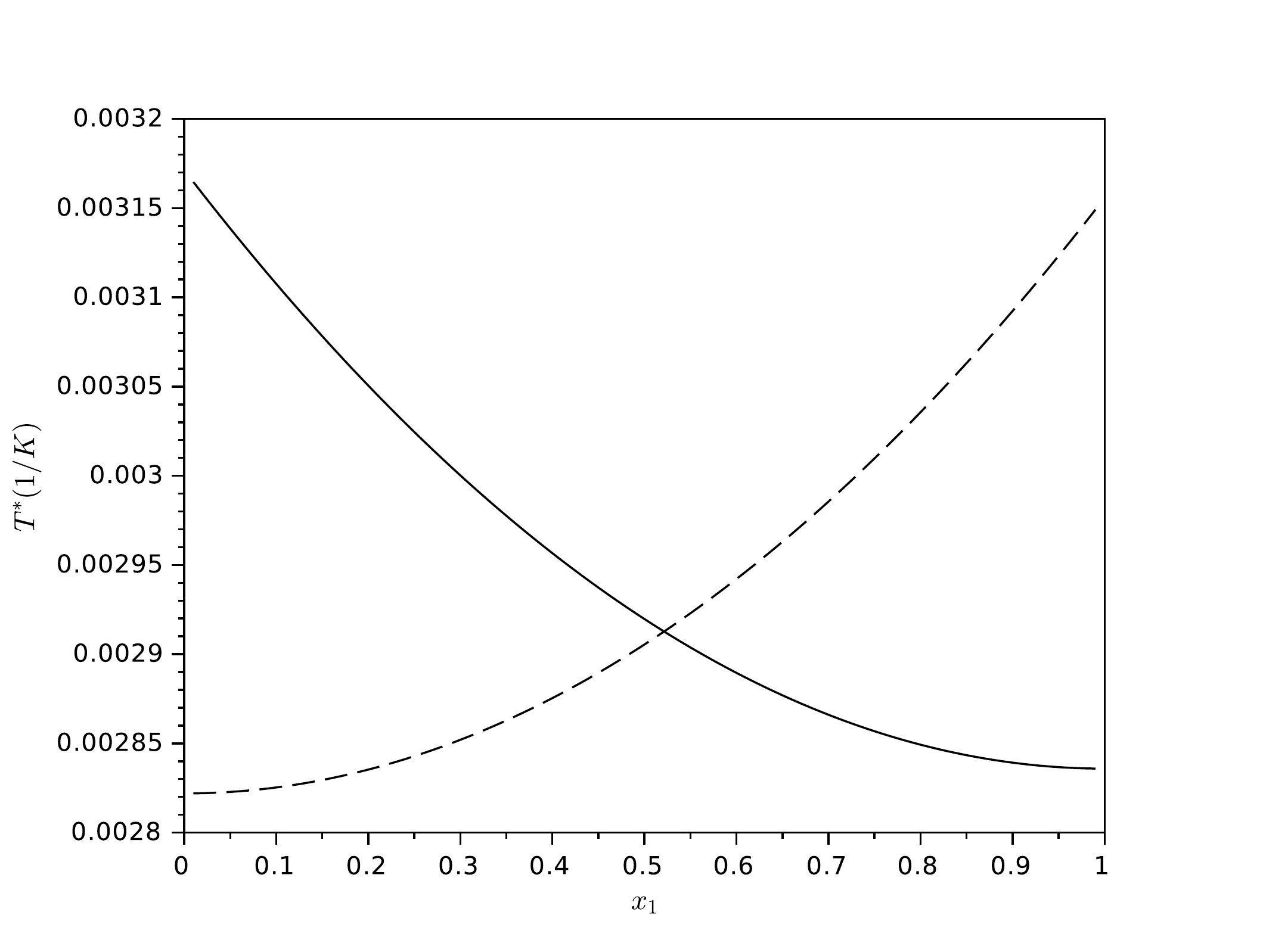}}
\label{fig:inter_parab_azeot}
\end{figure*}

Arranging the problem as the intersection of 
 two parabolas, Eqs.~(\ref{eq:azeotropy_log_final}), 
   the parabolas show minimum points at $x_1 = 0$ and $x_1 = 1$,
   due to the symmetrical nature of the activity coefficient model. The intersection of these two conicals represents the solution of the problem (azeotropic point). In fact, there are two intersections of the parabolas (the problem has two mathematical 
   solutions, but only one with physical significance, {\em i.e.}, with $0<x_1<1$). Figure \ref{fig:inter_parab_azeot} illustrates the intersection of the two curves in the azeotropic condition. The second intersection occurs at $x_1 = 407.86$, which, obviously, does not represent a physical solution.

\subsection{Problem 1B}

\begin{figure*}[t] \center 
\caption{Phase diagram for fictitious binary mixture with a double azeotrope at 500 mmHg.} \label{fig:phase_diagram_double_azeot}
\centerline{\includegraphics[width=0.9\textwidth]{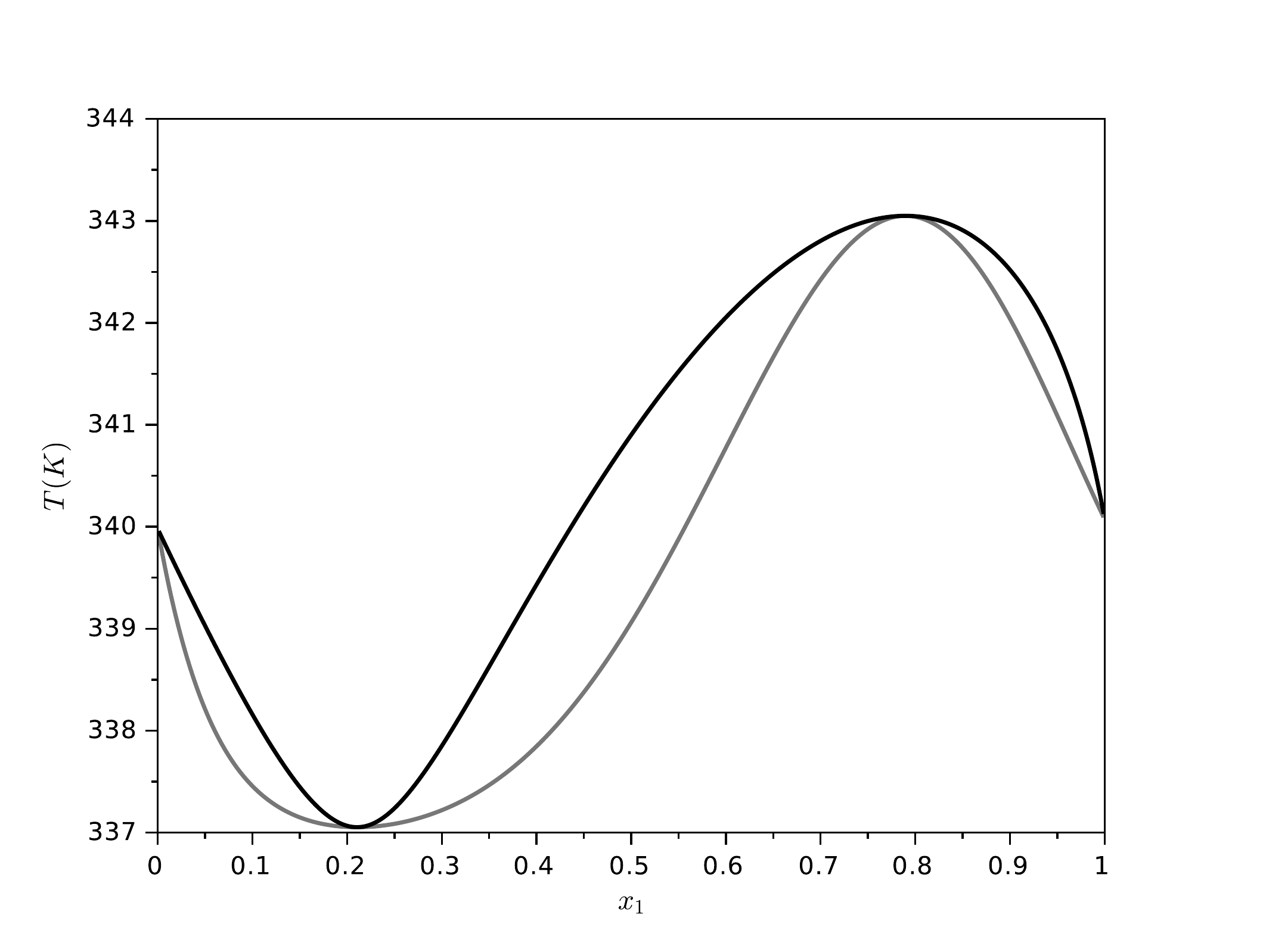}}
\end{figure*}

We use Problem 1B to illustrate the existence of a double azeotrope, even using this simplified approach. This rare phenomenon appears, for instance, in the binary mixture formed by hexafluorobenzene (1) + benzene (2). The normal boiling temperatures for hexafluorobenzene and benzene are, respectively, $T_{b,1}=353.4$ K and $T_{b,2}=353.3$ K. We also consider the coefficients for Margules model as $A_{12}=1$ and $A_{21}=-1$. Considering $B=17.2$, $C_{1}=-10.6(80.15+273.15)=-3744.45$ and $C_{2}=-10.6(80.25+273.15)=-3746.04$ the phase diagram was plotted and represented in Fig. \ref{fig:phase_diagram_double_azeot}, at 500 mmHg. It must be stressed that the values for the parameters $A_{12}$ and $A_{21}$ were not estimated using experimental data.
For this reason, we cannot consider that the binary mixture is, in fact, formed by hexafluorobenzene + benzene, but a fictitious mixture.
Nonetheless, these settings made it possible to produce a double azeotropy behavior. 


\begin{figure*}[t] \center 
\caption{The intersection of the graph of the two cubic polynomials in the physical domain (azeotropic condition).} \label{fig:inter_cubic_azeot}
\centerline{\includegraphics[width=0.9\textwidth]{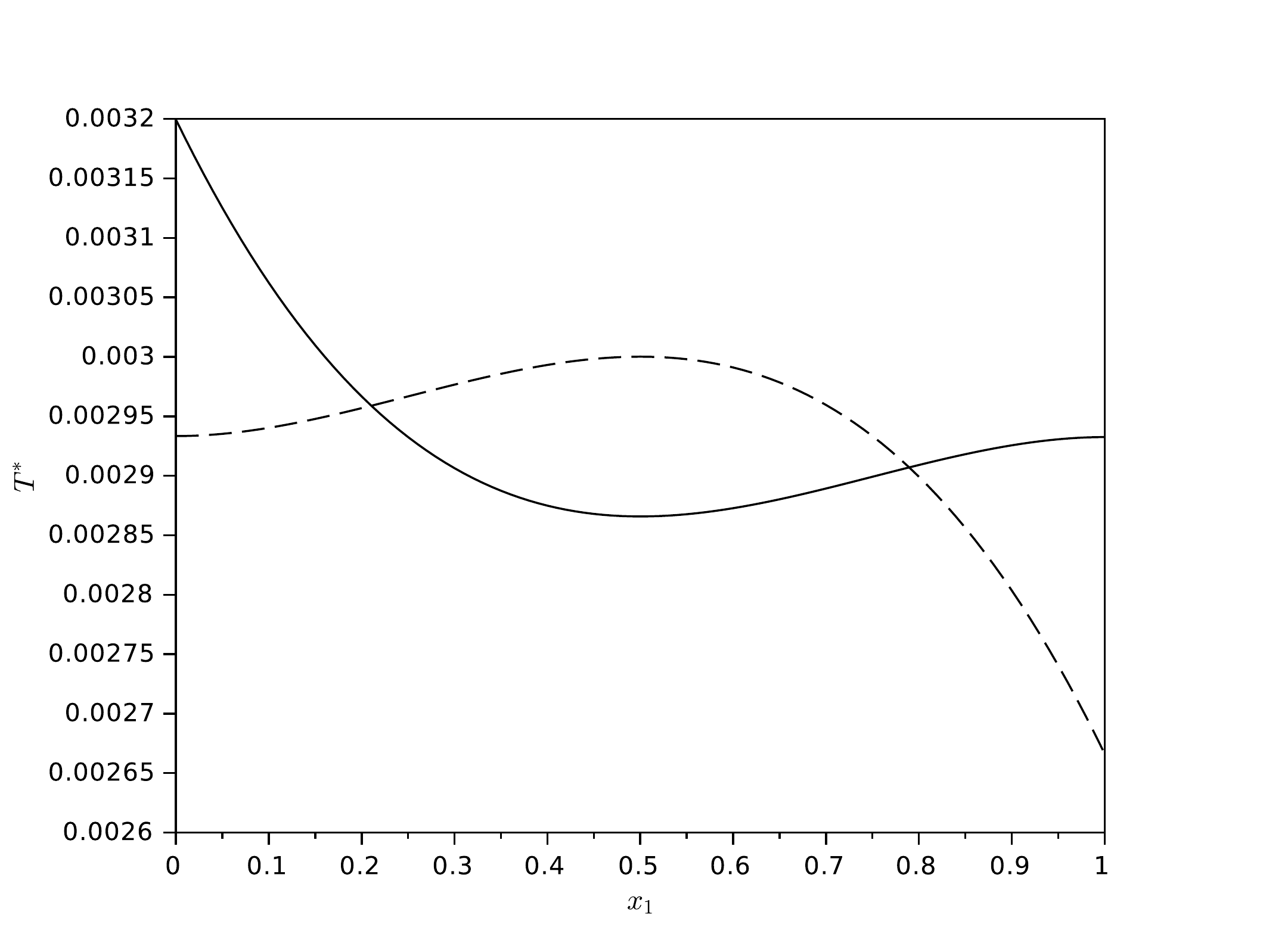}}
\end{figure*}

The azeotropic compositions, obtained by solving Eqs.~(\ref{19}), are the following: (i) $(x_1, T^{\ast})=(0.2104,0.00297)$; (ii) $(x_1, T^{\ast})=(0.7896,0.00292)$ and (iii) $(x_1, T^{\ast})=(-5299.75,1.59 \times 10^{8})$. The first two results represent physical solutions, and the third one is the intersection of the cubical curves outside the physical domain. In fact, we note that the inverse of the temperature in the third solution shows an extremely high value, indicating a temperature close to zero---without any physical significance, but useful to understand the vanishing of the double azeotrope, which will be explored subsequently. Figure \ref{fig:inter_cubic_azeot} illustrates the intersection of 
the two cubic curves, Eqs.~(\ref{22}), in the physically relevant part of the plane,
characterizing the azeotropic points.



\subsection{Problem 2A}

As pointed out in the description of the Formulation 2,
 Problem 2A, here we are interested in the effect of pressure $P$, or the parameter $B$, regarding the existence of the azeotropes.

\begin{figure*}[t] \center 
\caption{Effect of the pressure in the pre-images of the azeotropy problem. The variation of pressure does not lead to the approximation of the critical image. Thus, the chance of occurrence of a singularity is small.} \label{fig:effect_P_azeotrope_2A_amplification}
\centerline{\includegraphics[width=0.9\textwidth]{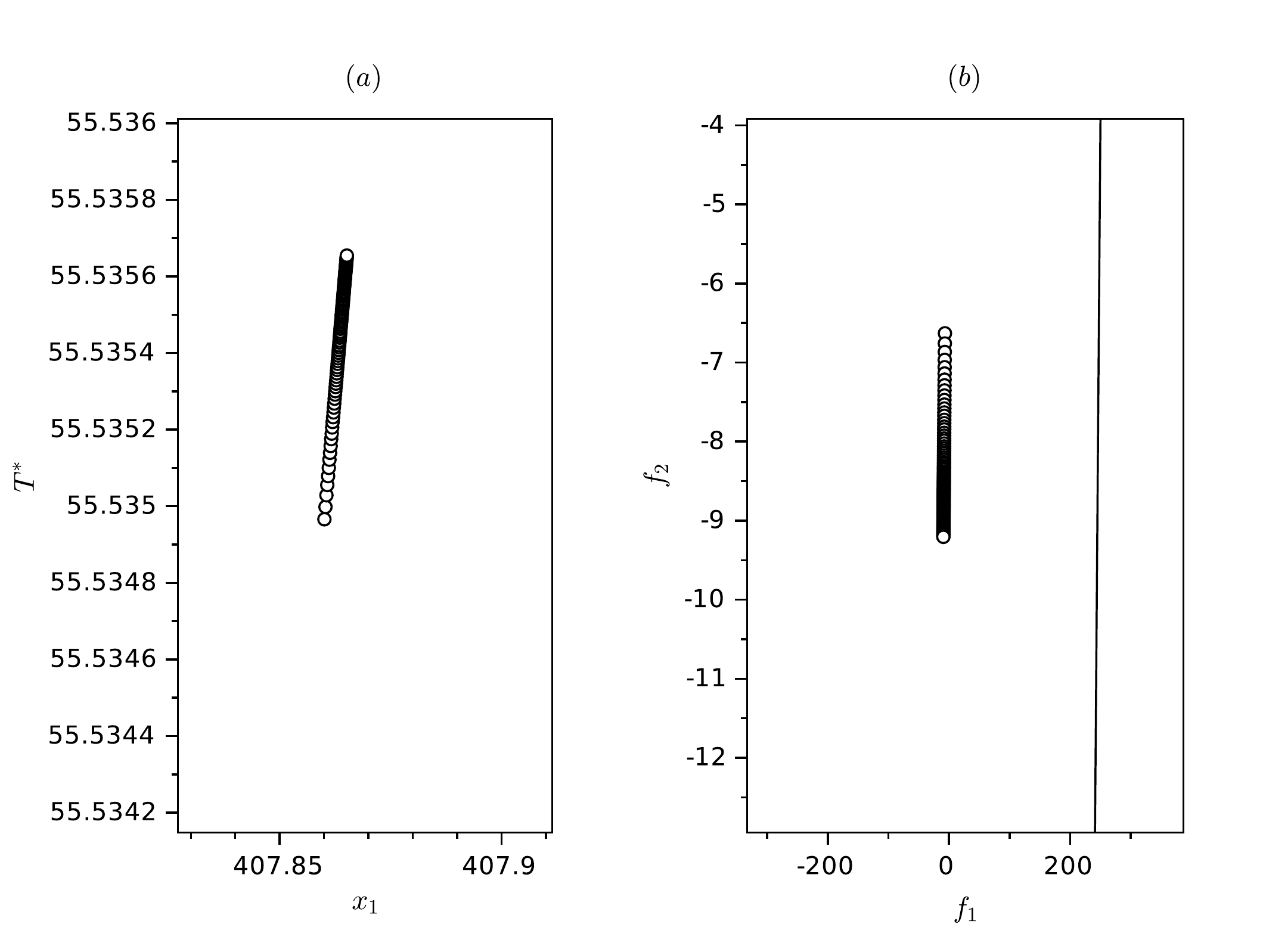}}
\end{figure*}

First, we analyze the effect of the pressure $P$. Figure \ref{fig:effect_P_azeotrope_2A_amplification} illustrates this behavior (we must note that, as pointed out previously, the critical curve is the same as found in Problem 1A and, in fact, it is far from the feasible domain). 

The effect of $P$ has little impact on the azeotropic coordinates, since the line that represents this variation is almost parallel to the critical image. We can conclude that variations in the pressure are not capable to induce the vanishing of the azeotrope, which would occur only if the critical curve was traversed.



\begin{figure*}[t] \center 
\caption{Effect of the parameter $B_1$ in the two pre-images of the azeotropy problem.} \label{fig:effect_B_azeotrope_2A}
\centerline{\includegraphics[width=0.9\textwidth]{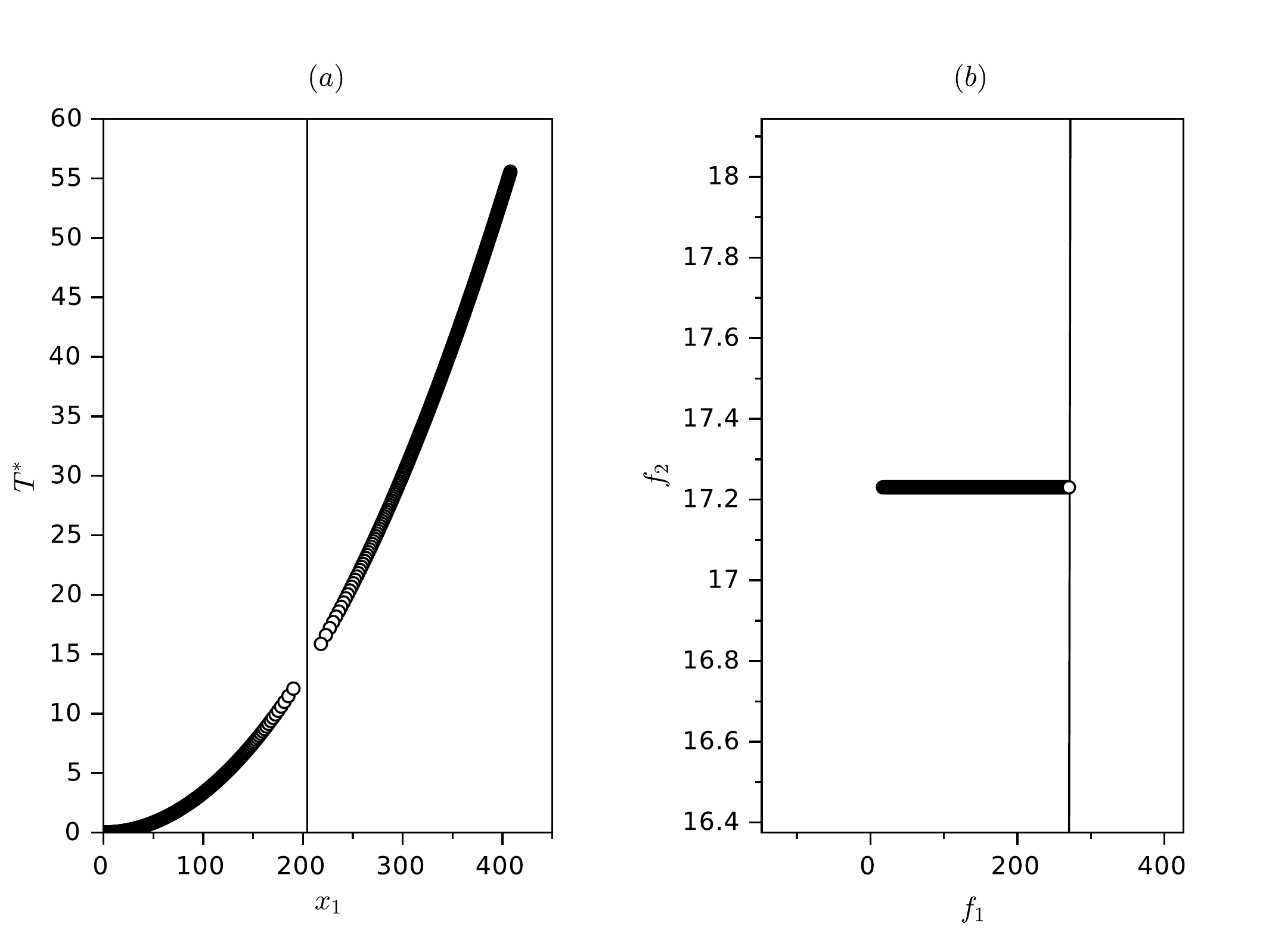}}
\end{figure*}

Figure \ref{fig:effect_B_azeotrope_2A} illustrates the pattern of the pre-images for a variation in the parameter $B_1$ from 17 to 270, with $B_2$ equals to 17.23 (the original value). We may note that when the critical image is approached, the two solutions of the nonlinear system tend to disappear, indicating---again---the existence of a fold point. Naturally, at some values of $B_1$ we obtain molar fraction greater than one (without physical sense). This is a consequence of the critical curve (a line parallel to the $T^{\ast}$ axis).


\subsection{Problem 2B}

\begin{figure*}[t] \center 
\caption{Effect of the pressure in the two pre-images of the azeotropy problem (Problem 2B).} \label{fig:effect_P_double_azeotrope_feasible}
\centerline{\includegraphics[width=0.9\textwidth]{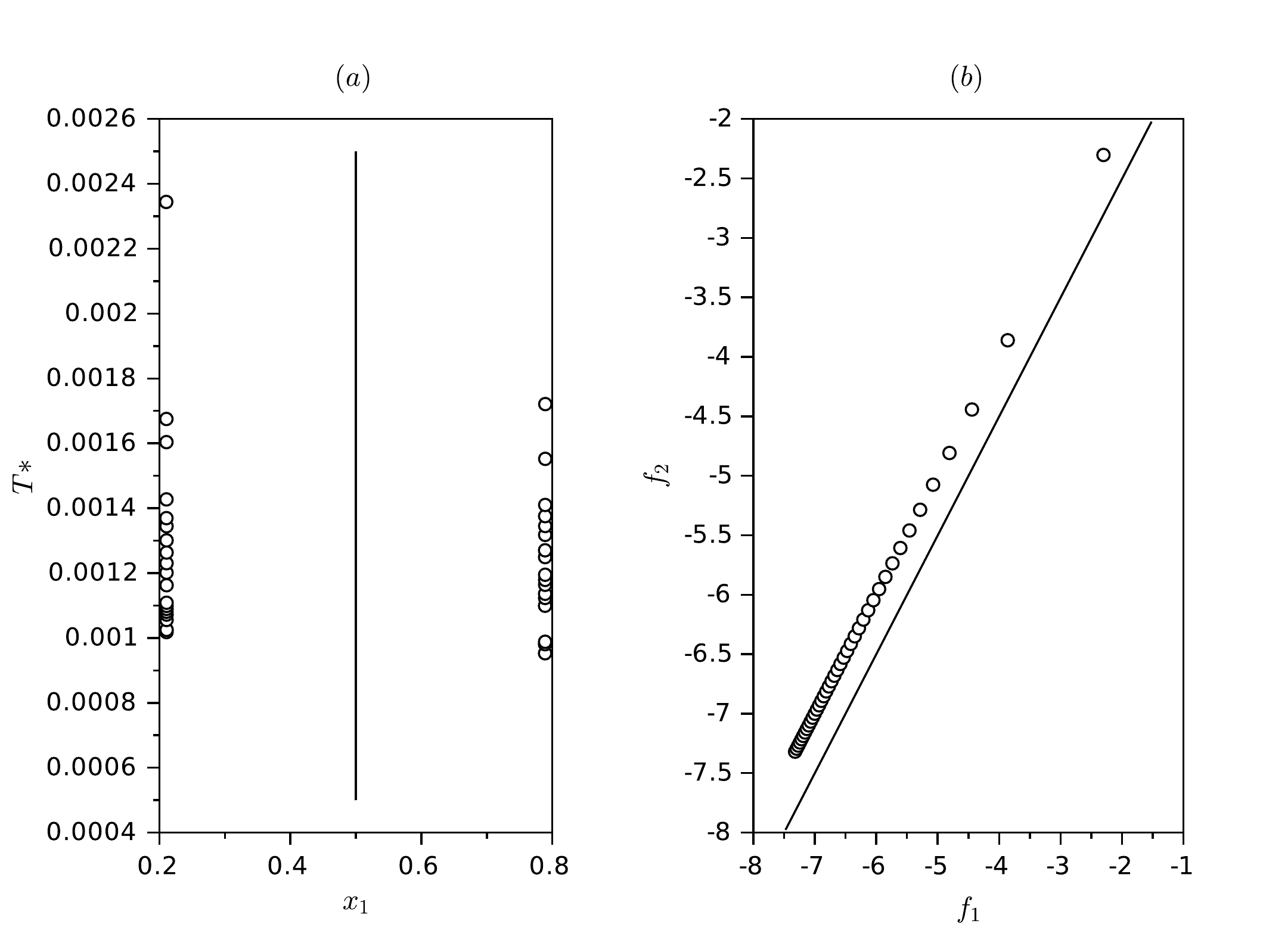}}
\end{figure*}

Again, we are interested in  the effect of 
variations of pressure $P$ and parameter $B$  in the existence of the azeotrope. With this modelling, the problem shows two critical curves, with $x_1 = 0.5$ (feasible) and $x_1 = -3533$ (unfeasible). Obviously, from the physical point of view, the unfeasible critical curve  is not interesting in the analysis. Then, we  focus in the feasible critical curve. Figure \ref{fig:effect_P_double_azeotrope_feasible} illustrates the effect of a pressure variation in the interval ranging from $10$ to $1510$ mmHg in the two feasible azeotropes ({\em i.e.}, with $0<x_1<1$). Clearly, the double azeotrope behavior does not suffer severe changes with the variation of the pressure. The same pattern is verified for the third solution (unfeasible, since $x_1>1$). It must be stressed that were are not particularly intended to verify if this particular variation of the pressure violates the assumption of ideal gas, considered in the model.


Finally, we will analyze the effect of these variations in $q = (B_1,B_2)$ for Problem 2B. We promote a variation in parameter $B_1$, maintaining a fixed value for $B_2$, with $B_2 = 17.23$. Again, we have two critical curves and two critical images. Then, it is interesting to assess the behavior (with respect of the number of solutions) in the vicinities of the two critical images, calculating the pre-images (the coordinates of the azeotropes).

\begin{figure*}[t] \center 
\caption{Effect of the parameter $B_1$ in the two pre-images of the azeotropy problem (Problem 2B).} \label{fig:effect_B1_double_azeotrope_feasible}
\centerline{\includegraphics[width=0.9\textwidth]{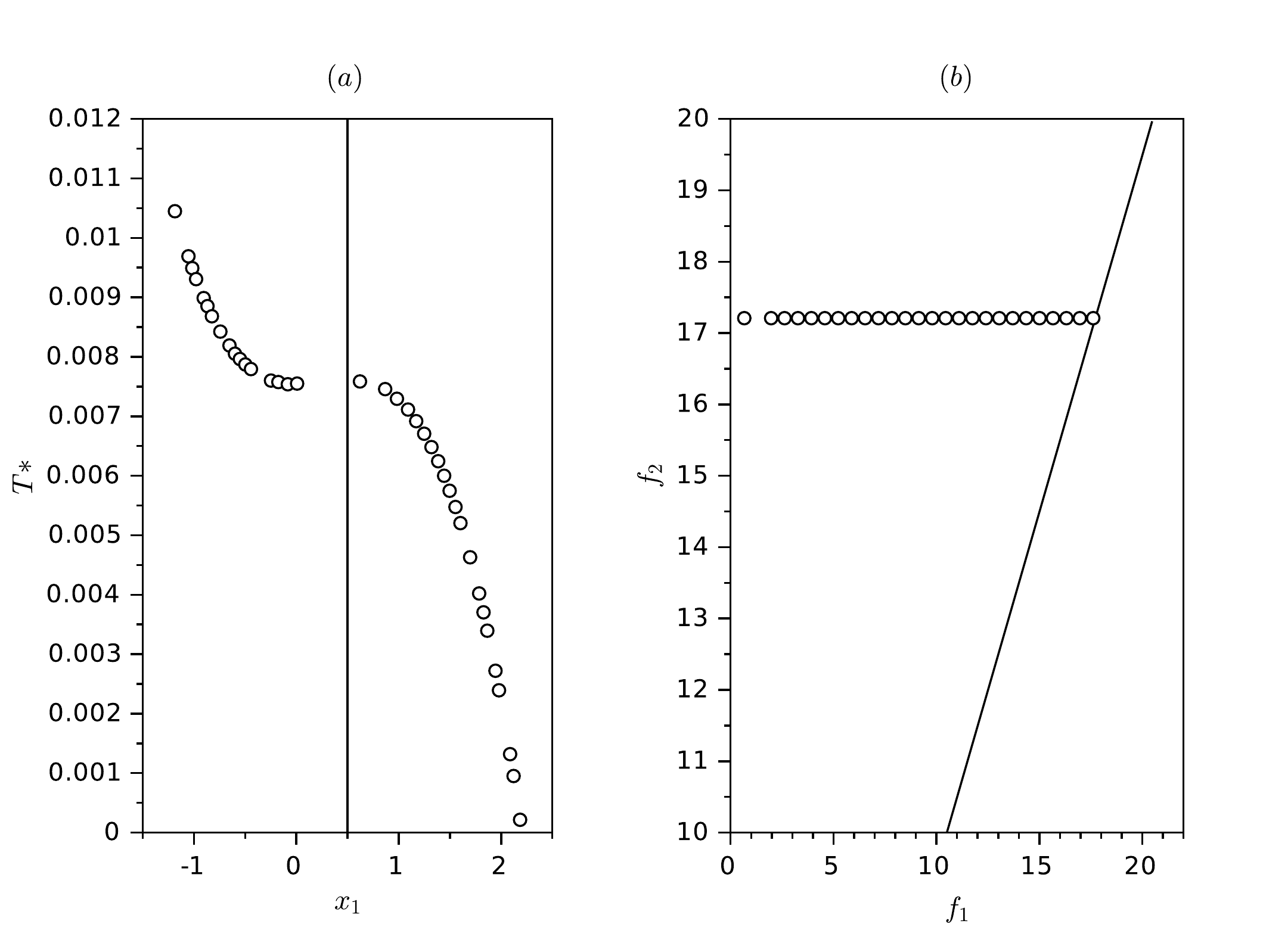}}
\end{figure*}

Figure \ref{fig:effect_B1_double_azeotrope_feasible} illustrates the effect of the parameter $B_1$ in the double azeotropy phenomenon. It must be stressed that a non-physical root appears, in this situation, with negative values for $x_1$. Thus, we can note that the double azeotrope disappear for $B_1$ close to $17.73$, where the two physical pre-images collapse.


\begin{figure*}[t] \center 
\caption{The branches of pre-images in the variation of the parameter $B_1$ (Problem 2B). Continuous red lines: critical curves and critical images. (a) Domain and (b) Codomain.} \label{fig:solution_branches}
\centerline{\includegraphics[width=0.9\textwidth]{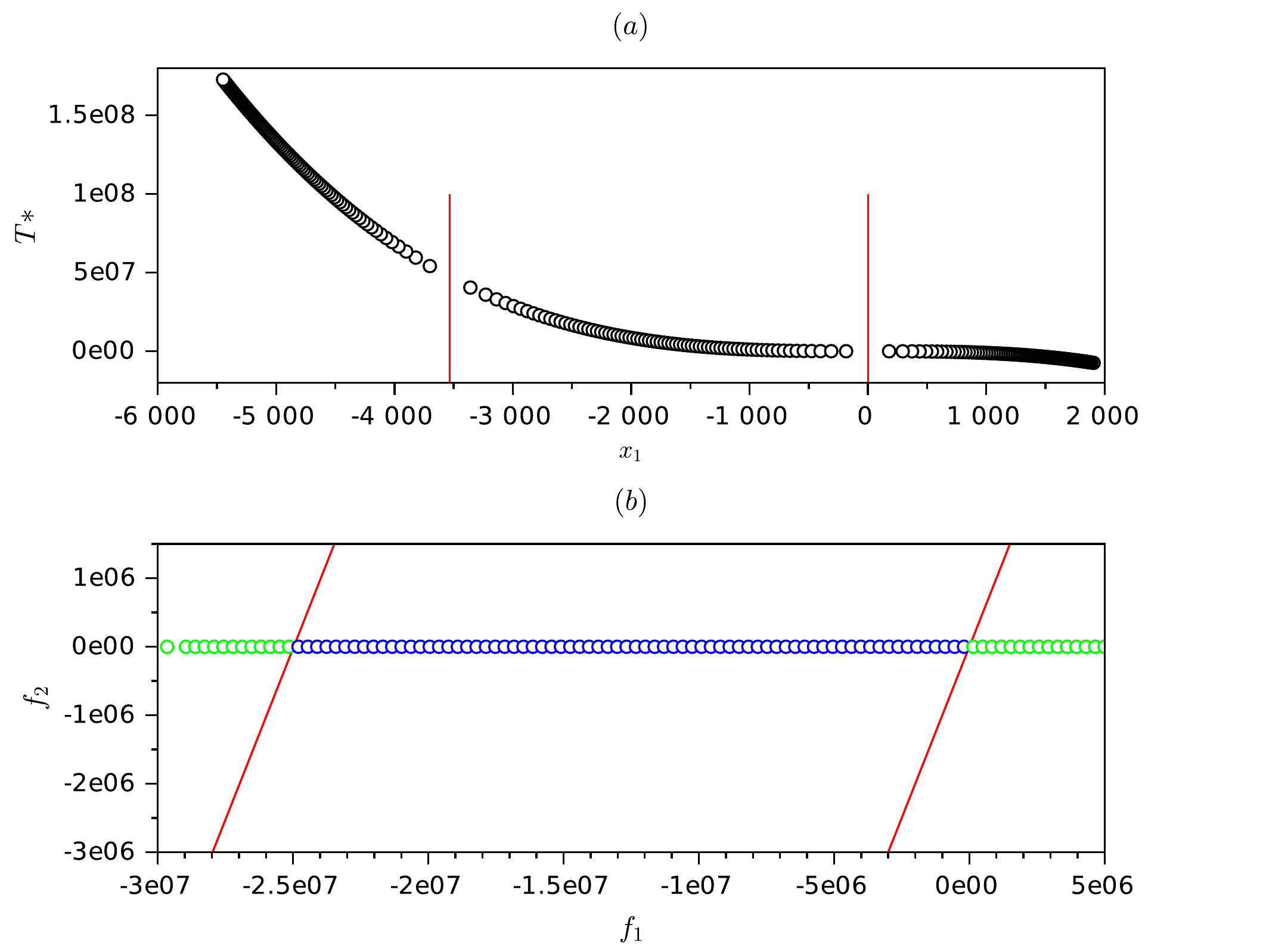}}
\end{figure*}

An interesting diagram is represented in Fig. \ref{fig:solution_branches}. This figure illustrates the behavior of the branches of pre-images for a large variation of the parameter $B_1$, in the range from $-5 \times 10^{7}$ to $5 \times 10^{6}$. For large negative values of $B_1$ only one solution is found---the branch for positive values of $x_1$ ($x_1$ close to 2000). This portion of the figure is marked in green points (in the image). As the value of $B_1$ approaches to the critical image, the value of $x_1$ of the unique solution is diminished. When the critical image is ``touched'', two pre-images ``are born'' with negative values of $x_1$. Now, three pre-images appear and the points are marked in blue. A further increase in the value of $B_1$, approaching to the second critical image brings on a movement of two pre-images to the physical domain, {\em i.e.}, $0<x_1<1$, and approaching to the second critical curve (with $x_1 = 0.5$). The third pre-image, with highly negative values, is moved far from the critical curve (with negative value for $x_1$). Close to the second critical curve ($x_1 = 0.5$), the double azeotropy phenomenon is verified (in fact, we have three pre-images, but only two in the physical range for $x_1$). When the second critical image is ``touched'', the two physical pre-images ({\em i.e.}, the coordinates of the double azeotrope) disappear. Beyond this point, only one (non-physical) solution persists (again, this portion of the diagram is marked with green points in the image).


\begin{figure*}[t] \center 
\caption{Phase diagram for fictitious binary mixture with a double azeotrope at 500 mmHg, with $B_1 = 17.73$ and $B_2 = 17.23$.} \label{fig:saddle_azeot}
\centerline{\includegraphics[width=0.9\textwidth]{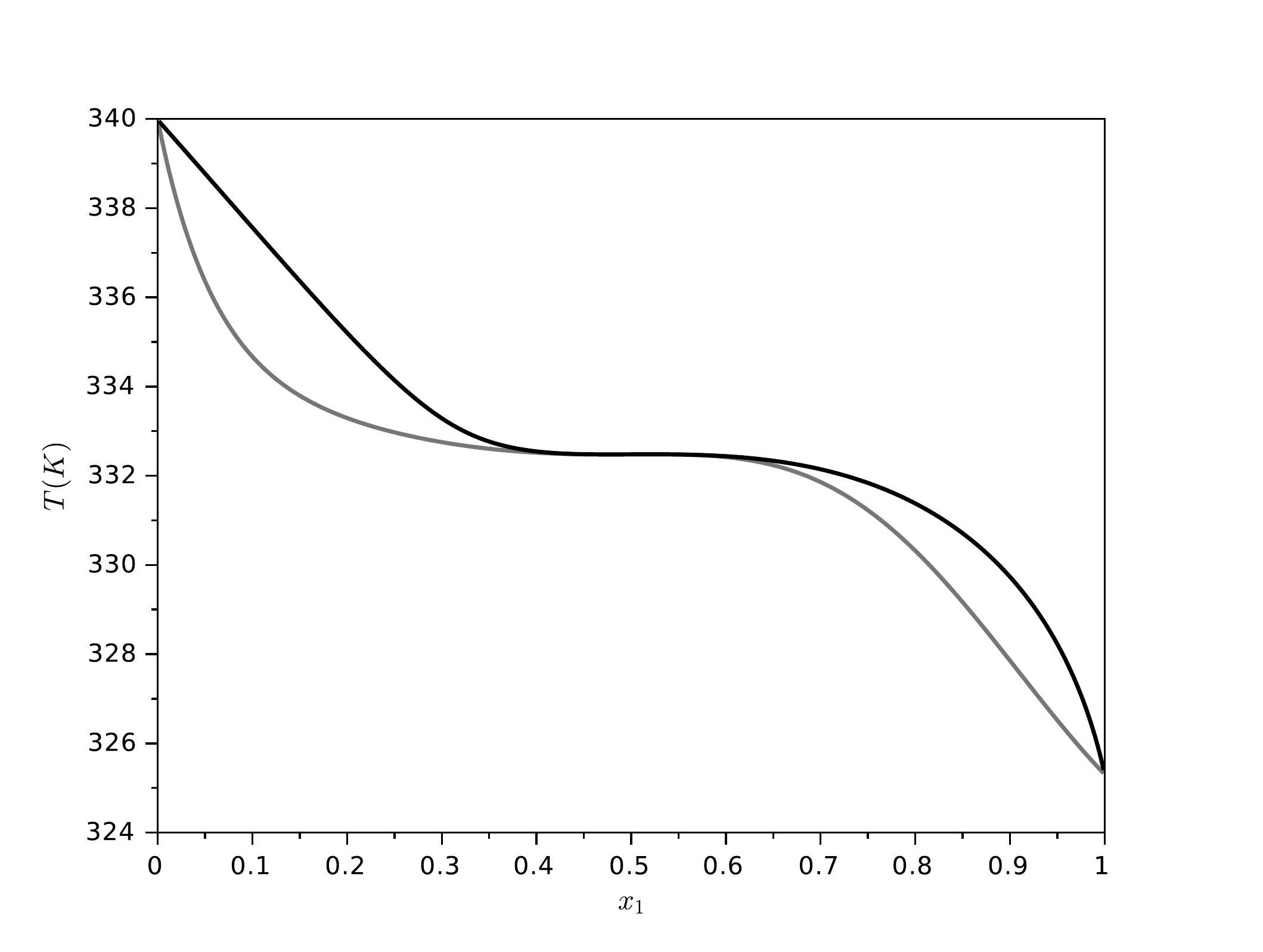}}
\end{figure*}

This behavior is consistent with a phase diagram at $P = 500$ mmHg, considering $B = (17.73, 17.23)$, as represented in Fig. \ref{fig:saddle_azeot}. We observe, for this value of the parameter $B$, that the double azeotrope is about to disappear. Once again, the phase diagram is obtained using a Newton-Raphson method for variables $x_1$ and $T$.
%

\section{Conclusions}

In this work we analyzed the azeotropy phenomenon from a geometric point of view, using simplified models for saturation pressures and activity coefficients. This kind of approach allows to understand some sophisticated thermodynamic concepts (including double azeotropy in binary mixtures) using not trivial but basic geometric concepts.

Two models and two different formulations for the nonlinear algebraic system that describes the calculation of an azeotrope were studied. The computational results illustrate that the azeotropy phenomenon and even double azeotropy can be viewed and predicted with simple mathematical and computational tools---for instance, as the intersection of conicals in the plane. Furthermore, the influence of some parameters---as system pressure and a parameter related to the vapor pressure of pure components---can be treated using the concept of functions from the plane to the plane.



\bibliographystyle{unsrt}
\bibliography{references.bib}

\end{document}